# Universal Minimum Heat Leak on Low-Temperature Metallic Electrical Leads

Alan M. Kadin

*HYPRES, Inc., 175 Clearbrook Road, Elmsford, NY 10523, USA*

**Abstract.** Low-temperature electronic systems require electrical leads which have low electrical resistance to provide bias current $I_{bias}$ without excessive voltage drop $V_{lead}$. But proper cryogenic design also requires high thermal resistance to maintain a minimal heat leak $\dot{Q}$ from the hot temperature $T_{hot}$ to the cold temperature $T_{cold}$. By the Wiedemann-Franz law, these requirements are in direct conflict, and the optimal configuration takes a particularly simple universal approximate form for the common case that $T_{cold} \ll T_{hot}$: $\dot{Q}_{min}/I_{bias} \approx V_{lead} \approx 3.6\, k_B T_{hot}/e$. This is applied here to the cryopackaging of RSFQ superconducting circuits on a 4 K cryocooler, but is equally applicable to other cryogenic systems such as superconducting sensor arrays at low and ultra-low temperatures.

**Keywords:** Cryogenic Design, Thermal Load, Current Leads, Cryocooler.

**PACS:** 72.15.-v, 85.25.-j, 44.10.+i, 07.20.Mc

## INTRODUCTION

In the design of cryogenic systems, it is important to minimize the heat flowing to the cold stage, in order to maintain thermal isolation and prevent excessive thermal loading. A critical path for heat is via metallic electrical leads that provide current bias $I_{bias}$ to superconducting and other cryogenic devices. The effective thermal resistance $R_{th}$ of the leads must be large, to prevent thermal conduction from the hot temperature $T_{hot}$ to the operating temperature $T_{cold}$. On the other hand, if the total $I_{bias}$ is large, the series electrical resistance $R_{el}$ of the leads must be small to prevent excessive Joule heating $I_{bias}^2 R_{el}$. In most cases, the leads consist of a normal metallic wire (Cu or Cu alloy), with both $R_{th}$ and $R_{el}$ dominated by electrons. Then according to the Wiedemann-Franz law (WF), $R_{th} \propto R_{el}$, and there is an optimum value that minimizes the thermal load $\dot{Q}$ on the cold stage. This is rather well known [1,2], and shown over many years in specific cryogenic systems. For example, we have developed optimized leads for a superconducting analog-to-digital converter (ADC), mounted on the 4 K stage of a two-stage cryocooler [3].

The focus of the present paper is not on this detailed design, but rather on the derivation of simple, general estimates for this optimized thermal load. In particular, we show that for the usual case that $T_{cold} \ll T_{hot}$, a simple analysis gives the approximate universal result that

$$\dot{Q}_{\min} / I_{bias} \approx 3.6 k_B T_{hot} / e. \qquad (1)$$

This result is perhaps not too surprising, and apart from the factor of 3.6 could be obtained directly from dimensional analysis. However, this does not seem to be generally appreciated in the low-temperature community, and may provide guidance for cryogenic design of high-current systems.

## UNIVERSAL HEAT LEAK RELATION

The WF law relates thermal and electrical conductivities by $\kappa = L\sigma T$, where $\kappa$ is the thermal conductivity, $\sigma$ the electrical conductivity, $L = (\pi k_B/e)^2/3 = 2.45 \times 10^{-8}$ W-$\Omega$/K$^2$ the Lorenz constant, $k_B$ is Boltzmann's constant, and $e$ the charge of the electron. This follows from the assumption that the effective thermal and electrical scattering times are the same, and holds reasonably well for most metals and alloys (superconductors being the major exception) over a wide range of T. Substantial deviations from WF can occur for pure metals at low T [2], but not as much for the metallic alloy wires that are typically used for cryogenic wiring. The WF law can also generally be applied directly to macroscopic properties at different temperatures, so that we have

$$R_{th} = R_{el} / LT^* \qquad (2)$$

where $T^* = (T_{hot} + T_{cold})/2$ is the average temperature for a wire of uniform cross section. This follows from $T^* = [\int(dx/\sigma)]/[L\int(dx/\kappa)] = (\int TdT)/(\int dT)$, since heat flow $\kappa\,dT/dx$ must be constant along the wire, neglecting Joule heating.

Now we can write an equation for the heat $\dot{Q}_1$ that is delivered to the cold stage on a single metallic lead:

$$\dot{Q}_1 = (T_{hot} - T_{cold})/R_{th} + I_{bias}^2 R_{el}/2 \\ = L(T_{hot}^2 - T_{cold}^2)/2R_{el} + I_{bias}^2 R_{el}/2 \quad (3)$$

Here we assume for simplicity that one-half of the Joule heat in the lead goes to the cold stage (and half to the hot stage). This is minimized by setting the two terms equal to each other (as can be shown by direct differentiation), which leads to an optimum $R_{el} = \sqrt{[L(T_{hot}^2 - T_{cold}^2)]}/I_{bias} \approx (\pi/\sqrt{3})k_B T_{hot}/eI_{bias}$, using here the common approximation that $T_{cold} \ll T_{hot}$. If we now consider the return lead as well (for a complete circuit), we obtain

$$V_{lead} = 2I_{bias} R_{el} \approx 2T_{hot}\sqrt{L} = \frac{2\pi}{\sqrt{3}} k_B T_{hot} \quad (4)$$

$$\dot{Q}_{min} = 2\dot{Q}_1 = I_{bias} V_{lead} \approx 3.6 I_{bias} k_B T_{hot}/e \quad (5)$$

These results are rather general and do not depend strongly on the specific assumptions.

## DISCUSSION AND CONCLUSIONS

Eqs. (4) and (5) apply to a variety of cryogenic systems, but we have focused on Nb-based RSFQ integrated circuits, designed to operate in liquid helium at 4.2 K. These consist of parallel-biased arrays of thousands of Josephson junctions, each biased just below its critical current ~ 0.2 mA, for a total bias current ~ 1 A. The bias voltage is ~ 2 mV, so the total power dissipated in the circuit is only ~ 2 mW. But the heat conducted down the bias leads is typically much greater. When immersed in liquid helium in the laboratory, thermal conduction on bias leads can generally be neglected. However, we have been packaging these systems [3] on two-stage commercial cryocoolers such as the Sumitomo SRDK-101D, with a specified heat lift of 5 W at the first stage at 60 K and 0.1 W at the second stage at 4 K. If we substitute 300 K into Eq. (5), we get $\dot{Q}_{min}/I_{bias}$ = 90 mW/A. This is clearly unacceptable for transmission to the 4 K cold stage, but can be easily accepted by the 60 K intermediate stage. If the leads are carefully thermalized on the 60 K stage, then the optimized heat load on the 4 K stage is reduced to ~ 15 mW/A, which is quite acceptable. These estimates are consistent with more detailed calculations and measurements for this system.

Similar issues are also present in other systems of superconducting devices. In particular, superconducting magnets require very large bias currents, which may be as large as 100 A or more. Such magnets are often operated in the persistent current mode, in which the leads can be disconnected, but the heat losses during ramping the fields up and down can be quite substantial. Another class of system consists of large arrays of superconducting imaging sensors, such as SQUIDs for magnetic fields and transition-edge sensors for photon detectors. These may operate at ultra-low temperatures (down to 0.1 K), where thermal conduction via leads is even more critical.

It is important to note that superconducting wires can strongly violate the WF law, and provide a way around the limitations of Eq. (5), since Cooper pairs can carry large electrical currents (with zero dissipation), but not thermal currents. However, this may require operation without the usual normal-metal stabilizer (Cu or Ag), which may complicate design and performance.

As an interesting aside, note that WF is also valid at high T, and the same set of arguments should also apply to the design of optimal thermal isolation in a high-T system, such as an electrically heated filament in a light bulb. For a tungsten filament, the temperature of operation should be ~ 3000 K. If we substitute this into Eq. (4), we obtain $V_{lead}$ = 0.9 V, even apart from the voltage drop across the filament itself. This is uncomfortably close to the 1.5 V supply voltage provided by a standard dry cell battery, and may explain why flashlights typically require at least two cells (3 V) for proper operation. A single-cell flashlight would operate marginally at best.

In conclusion, we have derived an approximate, but universal relation for the minimum heat load on the cold stage of a cryogenic system with a required current bias $I_{bias}$, $\dot{Q}_{min} \approx 2\pi k_B T_{hot} I_{bias}/\sqrt{3}$, and verified it for the particular case of an ADC chip on a cryocooler. This will assist in the design of superconducting systems with relatively large bias currents, such as RSFQ circuits and large detector arrays.

## ACKNOWLEDGMENTS

The author would like to thank Robert J. Webber, Deepnarayan Gupta, and others at HYPRES for helpful discussions. This research was supported in part by the US Office of Naval Research.